\documentclass[aps,pre,twocolumn,amsmath,amssymb,superscriptaddress,floatfix]{revtex4}
\usepackage{graphicx}
\usepackage{graphics}
\usepackage{bm}
\usepackage[usenames]{color}
\usepackage{colordvi}
\usepackage{units}
\usepackage{bbm}
\usepackage{enumitem}
\usepackage{calc}

\usepackage[utf8]{inputenc}
\usepackage{amsmath}
\usepackage{amssymb}
\usepackage{epsfig}
\usepackage{rotating}

\usepackage{algorithm}

\usepackage{algpseudocode}

\usepackage{amsfonts}

\newcommand{\nnn}{\nonumber\\}

\newcommand{\mb}[1]{\mathbf{#1}}

\newcommand{\pdou}[2]{\frac{\partial {#1}}{\partial {#2}}}
\newcommand{\inuov}[3]{\int_{#1}^{#2} d{#3}\,\,}
\newcommand{\ablfac}[2]{\left(1-\frac{k({#1},{#2})}{k_m(#2)}\right)}
\newcommand{\tm}{t_{\text{max}}}

\newcommand{\vrtubrntn}{\left(\mb{r},t|\mb{r}_0,t_0\right)}

\newcommand{\randxtxn}[3]{\frac{\nabla\left(P_D({#1}, 
{#2}|{#3})\right)\cdot \mathbf{n}_{#1}}{\int_{\partial G} dF\,\, \nabla 
\left( P_D(\mathbf{r}, {#2}|{#3})\right)\cdot \mathbf{n}_{\mathbf{r}} }}
\newcommand{\internxtxn}[3]{\frac{P_D(#1,#2|#3)}{\int_G d\mb{r'} 
P_D(\mb{r}', #2 |#3)}}

\newcommand{\non}{\nonumber}

\begin{document}

\title{Efficient kinetic Monte Carlo method for reaction-diffusion 
processes\\
with spatially varying annihilation rates}

\author{Karsten Schwarz}
\email{kschwarz@lusi.uni-sb.de}
\author{Heiko Rieger}
\email{h.rieger@mx.uni-saarland.de}
\affiliation{Theoretische Physik, Universit\"at des Saarlandes, 66041
Saarbr\"ucken, Germany}
\date{\today}

\begin{abstract}
We present an efficient Monte Carlo method to simulate
reaction-diffusion processes with spatially varying particle
annihilation or transformation rates as it occurs for instance in the
context of motor-driven intracellular transport. Like Green's function
reaction dynamics and first-passage time methods, our algorithm
avoids small diffusive hops by propagating sufficiently distant
particles in large hops to the boundaries of protective domains.
Since for spatially varying annihilation or transformation
rates the single particle diffusion propagator is not known analytically,
we present an algorithm that generates efficiently either
particle displacements or annihilations with the correct statistics,
as we prove rigorously. The numerical efficiency of the algorithm is
demonstrated with an illustrative example.\\
{\bf Keywords:} computer simulations, kinetic Monte Carlo, 
reaction-diffusion processes

\end{abstract}

\maketitle

\section{Introduction}

Kinetic Monte Carlo simulations are frequently used in
various fields to analyze the spatio-temporal evolution of
systems consisting of many freely diffusing particles that
can collide, react, transform or annihilate. Spatial as well
as stochastic aspects are important when diffusion is not
sufficiently fast to make the system well-stirred and the
number of reactants within diffusion range is small. In
this case a mean-field description, for instance with a set
of coupled reaction-diffusion equations, is inappropriate.
Moreover, in the limit of extreme dilution methods using
a discretization of the underlying stochastic reaction-diffusion
system, either in time \cite{bray} or in space \cite{elf1,elf2},
become computationally inefficient.

The currently most efficient methods to simulate extremely diluted
reaction-diffusion systems are Green's function reaction dynamics
\cite{tenwolde1,tenwolde2} and first-passage kinetic Monte Carlo
methods \cite{oppelstrup1,oppelstrup2,donev}. In essence they avoid
the small diffusion hops of a conventional random walk or Brownian
dynamics simulation by propagating particles over long distances
through a sequence of large displacements. The latter are
generated stochastically according to the exactly known Green's
function for a freely diffusing particle within so-called protective
domains that are free from other particles. The typical size of these
protective domains is inversely proportional to the particle density
and the larger these domains are (i.e. the smaller the particle density is)
the more efficient the algorithm is.

In general, during the free diffusion the particle can
also be annihilated or transformed with a rate $k$ into a
different species, in which case the Green's function is
still exactly know. In this paper we address the question
how to propagate the particles when the annihilation rate
varies in space and time, denoted as $k(\mb{r},t)$. This problem
arises for instance in the context of motor-driven
intracellular transport, where particles (or cargos) can in addition
to diffusion and reaction also attach to a cytoskeleton filament
and move ballistically with a constant speed in
the direction of the filament. A continuum description
of the diffusive and ballistic modes of motion \cite{loverdo,loverdo2}
involves the filament density $\rho(\mb{r},t)$ which determines the
local rate with which freely diffusing particles make a
transition into the ballistic state. In a typical cell the filament
density is spatially inhomogeneous and thus has to
be taken into account during the propagation of particles
on large scales. Analogous examples arise in systems in
which the annihilation of particles depends on a spatially
inhomogeneous concentration field of an abundant reaction
partner (i.e. whose density is much larger such that
a continuum description is appropriate for it).

Green's function reaction dynamics and first-passage time
Monte Carlo methods reduce the simulation of a many-particle
reaction-diffusion system to individual particles that
diffuse freely as long as other particles are sufficiently distant
(i.e.\ outside the interaction range), and perform a reaction event
once a particle pair reaches a minimum distance. Algorithmically
one can ensure free diffusion for instance by estimating the
maximum diffusion distance \cite{tenwolde1,tenwolde2} until a reaction is
scheduled or by the definition of protective domains for each particle
\cite{oppelstrup1,oppelstrup2,donev} depending on the actual
arrangement of neighboring particles. In both cases one then
utilizes the free diffusion propagator within predefined
domains to generate stochastically a time when either
the maximum distance is underrun or a protective domain boundary is reached.
For free diffusion this is achieved using the analytically
known Green's function, but for free diffusion with spatially varying
annihilation rates this propagator is unfortunately not analytically
available.

Thus in this paper we consider a freely diffusing single
particle in an arbitrary domain $G\in R^n$ that can be annihilated
with a time and space dependent rate $k(\mb{r},t)$.
In general, annihilation means a transition into a different
species that is not considered in the present reduced
setup. For a particle initially at time $t_0$ located at $\mb{r_0}\in G$
this diffusion-annihilation process is described by the following
diffusion-annihilation equation

\begin{eqnarray}
\pdou{P(\mb{r},t|\mb{r}_0,t_0)}{t}=D\Delta P(\mb{r},t|\mb{r}_0,t_0)
- k\left(\mathbf{r},t\right) P(\mb{r},t|\mb{r}_0,t_0)\;,\nonumber\\
\quad
\label{Dglallgemein}
\end{eqnarray}
where $P(\mb{r},t|\mb{r}_0,t_0)$ is the probability density to find
the particle at time $t$ at $\mb{r}\in G$. For arbitrary $k(\mb{r},t)$
and arbitrary $G$ there is no analytic solution of
Eq. (\ref{Dglallgemein}) available. In principle this equation can be
solved numerically, but in the context of a general reaction-diffusion
system (involving many particles and several particle species) using
for instance the first-passage Monte Carlo method this is unfeasible:
Here one needs for each particle hop the whole first-passage time
distribution for a particle to reach the protective domain boundary
$\partial G$, which is computationally too demanding to be carried out
in the innermost loop of the algorithm.

Therefore we present in this paper an algorithm that samples times
$t > t_0$ and positions $\mb{r}$ for arbitrary annihilation rates 
$k(\mb{r},t)$ and arbitrary domains for which a particle diffusing according
to Eq. (\ref{Dglallgemein}) either a) reaches the boundary for the first time
($\mb{r}\in \partial G$) or b) is annihilated ($\mb{r}\in G$). In
addition, a slightly modified version of the algorithm generates
the whole probability density $P(\mb{r},t|\mb{r}_0,t_0)$ within $G$,
meaning it solves Eq. (\ref{Dglallgemein}) stochastically.

The paper is organized as follows: Section \ref{basic_KMC} defines all
probability densities and flows used throughout this paper. Based on
the ideas of \cite{tenwolde1,tenwolde2,oppelstrup1,
oppelstrup2,donev}, section \ref{hom_rate} presents an algorithm for
the sampling of $(\mb{r},t)$ on arbitrary domains $G$ in the case of a
spatially homogeneous but temporally varying annihilation rate
$k(\mb{r},t)=k(t)$. Section \ref{inhoratsec} generalizes this method
to spatially inhomogeneous rates $k\left(\mathbf{r},t\right)$,
proves its correctness and discusses its efficiency. Finally section
\ref{simuerg} shows an application example of this method.

\section{Definitions}

\label{basic_KMC}

In this section the probability densities and flows used later on are 
defined.
Let $P(\mb{r},t|\mb{r}_0,t_0)$ be the probability density solving the 
diffusion-annihilation
equation (\ref{Dglallgemein}) within the domain $G$ with boundary 
$\partial G$, possibly
partly absorbing, partly reflecting. The particle annihilation generates 
a probability flow
$f_a(\mb{r},t|\mb{r}_0,t_0)$ out of the system given by

\begin{eqnarray}
f_a(\mb{r},t|\mb{r}_0,t_0)=k(\mb{r},t)\cdot P(\mb{r},t|\mb{r}_0,t_0)\;.
\label{strom}
\end{eqnarray}

The probability flow $f_b(\mb{r},t|\mb{r}_0,t_0)$ at the absorbing parts 
of the boundary at time $t$ at position
$\mb{r}\in\partial G$ is given by

\begin{eqnarray}
f_b(\mb{r},t|\mb{r}_0,t_0)=-D\,\,\nabla P(\mb{r},t|\mb{r}_0,t_0)\cdot 
\mb{n}_{\mb{r}} \;, \label{randabsorb}
\end{eqnarray}

where $\mb{n}_\mb{r}$ denotes the outward pointing unity vector
perpendicular to the boundary $\partial G$ at $\mb{r}$. Consequently
$P(\mb{r},t|\mb{r}_0,t_0)$ is not normalized for $t>t_0$.
The corresponding probability density $\rho_e (t|\mb{r}_0,t_0)$ for
an annihilation or absorption event is given by

\begin{eqnarray}
\rho_e (t|\mb{r}_0,t_0)&=&-\frac{d}{dt}\left[\int_G d\mb{r}\,\, 
P(\mb{r},t|\mb{r}_0,t_0)\right]\nnn &=&\alpha(t|\mb{r}_0,t_0)+ 
\beta(t|\mb{r}_0,t_0)
\end{eqnarray}
\begin{eqnarray}
\text{with }\quad \alpha(t|\mb{r}_0,t_0)&=&\int_G d\mb{r}\,\, 
f_a(\mb{r},t|\mb{r}_0,t_0)\nnn \text{and }\quad 
\beta(t|\mb{r}_0,t_0)&=&\int_{\partial G} dF\,\, 
f_b(\mb{r},t|\mb{r}_0,t_0)\;,\nonumber
\end{eqnarray}
where $dF$ denotes the surface element at position 
$\mb{r}\in\partial G$.
Hence, the task is to sample the pairs $(\mb{r},t)$ in statistical 
agreement to
$f_a(\mb{r},t|\mb{r}_0,t_0)$ and $f_b(\mb{r},t|\mb{r}_0,t_0)$, i.e the 
statistic of $t$ will be according to $\rho_e$.

In the following we also need the probability distribution of a freely 
diffusing particle
$P_D\vrtubrntn$ without annihilation, which obeys

\begin{eqnarray}
\pdou{P_D\vrtubrntn}{t}=D\,\,\Delta P_D\vrtubrntn\;. \label{Dglnurdiff}
\end{eqnarray}

The probability density for being absorbed at the boundary for a purely 
diffusing particle at time $t$ is given by
\begin{eqnarray}
\rho^D_{b}(t|\mb{r_0},t_0)=-\frac{d}{dt}\int_G d\mb{r}\,\,P_D\vrtubrntn 
\label{absorb_dens}
\end{eqnarray}
and the probability density of the absorbing position $\mb{r}\in\partial 
G$ under the condition that the absorption takes place at time t is given by
\begin{eqnarray}
\rho^D_f(\mb{r}|t,\mb{r}_0,t_0)=\frac{\nabla P_D\vrtubrntn \cdot 
\mb{n}_{\mb{r}}}{\int_{\partial G} dF\,\,\, \nabla P_D\vrtubrntn 
\cdot \mb{n}_{\mb{r}}}\;.
\end{eqnarray}
Using the Gauss's theorem and Eq. (\ref{Dglnurdiff}) in the denominator, 
one obtains:
\begin{eqnarray}
\rho^D_f(\mb{r}|t,\mb{r}_0,t_0)=-D \frac{ \nabla P_D\vrtubrntn \cdot 
\mb{n}_{\mb{r}} }{\rho^D_{b}(t|\mb{r_0},t_0)}\;.
\end{eqnarray}

For spatially homogeneous annihilation rates $k(\mb{r},t)=k(t)$ the 
annihilation
process decouples from all spatial variables, i.e. the solution of
Eq. (\ref{Dglallgemein}) can be written as

\begin{eqnarray}
P\vrtubrntn=e^{-\int_{t_0}^t k(t')dt'}\cdot 
P_D\vrtubrntn\;.\label{sol_t_dep_rate}
\end{eqnarray}
Hence, the probability density of being annihilated at time $t$ under 
the condition of not being absorbed at the boundary before for a 
spatially homogeneous rate $k(t)$ is given by
\begin{eqnarray}
\rho^D_{a}(t|t_0)=-\frac{d}{dt} \left[ e^{-\int_{t_0}^t 
k(t')dt'}\right] \label{rate_dens}
\end{eqnarray}
and the probability density of the annihilation position $\mb{r}\in G$ 
under the condition that the particle is annihilated at time $t$ is given by
\begin{eqnarray}
\rho^D_n(\mb{r}|t,\mb{r_0},t_0)=\frac{P_D(\mb{r},t|\mb{r}_0,t_0)}{\int_G 
d\mb{r}'\,\, P_D(\mb{r}',t|\mb{r}_0,t_0)}\;,
\end{eqnarray}
which is equal to the probability density of a purely diffusing particle under the condition of 
not being absorbed.

\section{Homogeneous annihilation rate}
\label{hom_rate}

In this section we present an algorithm that samples times
$t > t_0$ and positions $\mb{r}$  for homogeneous annihilation
rates $k(\mb{r},t)=k(t)$ and arbitrary domains $G$ for which a 
particle diffusing according to Eq. (\ref{Dglallgemein}) either a) 
reaches the boundary for the first time
($\mb{r}\in \partial G$) or b) is annihilated ($\mb{r}\in G$).

Assuming the ability to generate random numbers according to all the
densities, which were defined in the previous section, a correct way
of sampling $(\mb{r},t)$ is shown in Algorithm \ref{alg1}:

\begin{algorithm}
\caption{ KMC 1}
\label{alg1}
\begin{algorithmic}[0]
\State{{\bf{Input}}: $\mathbf{r}_0$, $t_0$}
\State{\bf{Output}: $\mb{r}, t$}
\State $t_a\gets$ random number according to $\rho^D_{a}(\cdot|t_0)$
\State $t_b\gets$ random number according to 
$\rho^D_b(\cdot|\mathbf{r}_0,t_0)$
\State $t\gets\text{min}(t_a,t_b)$
\If{($t_a<t_b$)}
\State $\mathbf{r}\gets$ random position according to 
$\rho^D_n(\cdot|t_a,\mb{r}_0,t_0)$

\Else
\State $\mathbf{r}\gets$ random position at the boundary $\partial G$
\State $\quad\quad\,$ according to $\rho^D_f(\cdot|t_b,\mb{r}_0,t_0)$

\EndIf
\State\Return ($\mb{r},t$)
\end{algorithmic}
\end{algorithm}

The probability density $A(\mb{r},t)$ that the algorithm produces an
annihilation at time $t$ at position $\mb{r}$ is then given by

\begin{eqnarray}
A(\mb{r},t)&=&\rho_a^D(t|t_0) \left(\int_t^\infty dt_b\,\, 
\rho_b^D(t_b|\mb{r}_0,t_0)\right) \rho^D_n(\mb{r}|t,\mb{r_0},t_0)\nnn
&=&\rho_a^D(t|t_0)\,P_D\vrtubrntn\nnn
&=&f_a\vrtubrntn\;.\non
\end{eqnarray}

The probability density $B(\mb{r},t)$ that the algorithm delivers an
absorption at time $t$ at position $\mb{r}\in \partial G$ is given by

\begin{eqnarray}
B(\mb{r},t)&=&\left(\int_t^\infty dt_a\,\, \rho_a^D(t_a|t_0)\right) 
\rho_b^D(t|\mb{r}_0,t_0)\,\rho^D_f(\mb{r}|t,\mb{r_0},t_0)\nnn
&=&f_b\vrtubrntn\;.\non
\end{eqnarray}

Consequently, the statistic of random pairs $(\mb{r},t)$ generated
in this way coincides with $f_a(\mb{r},t|\mb{r}_0,t_0)$ and
$f_b(\mb{r},t|\mb{r}_0,t_0)$ and is therefore correct.

One problem remains: As there are no analytic
solutions for Eq. (\ref{Dglnurdiff}) available for arbitrary domains
$G$, it is not possible to sample the quantities $\rho_f^D$, $\rho_n^D$
and $\rho_b^D$ directly for arbitrary domains $G$ and arbitrary
boundary conditions. Only a direct sampling of $\rho_a^D$ is possible,
as there is always an analytic expression for the corresponding
distribution function available: 
\begin{eqnarray}
F_a^D(t|t_0)=1- e^{-\int_{t_0}^t k(t')dt'}\,\,.
\end{eqnarray}
That means, Algorithm \ref{alg1} is useful only in some special
geometries $G$.

Nevertheless, it is possible to use these special cases to sample the
random pair $(\mb{r},t)$ for arbitrary domains. Two different methods
will be shown now.

\subsection{Subset method}

In \cite{oppelstrup1, oppelstrup2,donev}, a kinetic Monte Carlo method
for the simulation of reaction-diffusion processes of
many-body systems is presented. It is based on the fact that
there are analytic solutions of Eq. (\ref{Dglnurdiff}) for
some simple domains $G'$ and boundary conditions. The appendix shows a
list with some of these domains in one, two and three dimensions and
derives expressions for distribution functions, which are necessary
for the usage of the inversion method \cite{devroye}.\\

If $G'$ denotes a subset of $G$ with $\mb{r}_0\in G'$, the shape of
$G\setminus G'$ will not matter for the particle, as long as the
particle has not left $G'$ for the first time. Hence, we can restrict
the description of the particle's motion to $G'$ until it leaves $G'$
for the first time. Mathematically we are dealing with a
first-passage-problem in $G'$. Its solution is given by
Eq. (\ref{Dglnurdiff}) on $G'$ according to absorbing boundary
conditions at the interior of $G$ and the boundary conditions of $G$
at common boundaries of $G$ and $G'$ (as far as they exist).

Assuming that we are able to sample all occurring densities, a random
pair $(\mb{r},t)$ for $G'$ can be generated, as shown in Algorithm
\ref{alg1}. If annihilation takes place ($t_a<t_b$), the particle is
annihilated before it leaves $G'$ and therefore it is not influenced by
the restriction to $G'$. If the particle reaches the boundary of $G'$
($t_b<t_a$), two possibilities have to be distinguished: For $\mb{r}\in
\partial G$ it reaches an absorbing boundary of $G$ and the algorithm
will stop. Otherwise, the particle continues its diffusive motion
under the condition of having been at position $\mb{r}_0=\mb{r}$ at
time $t_0=t$. As it is always possible (see appendix) for an arbitrary
$\mb{r}_0$ to find a subset of $G$ where there are all needed analytic
expressions available, we can go on this way until the particle is
annihilated or absorbed at the boundary of $G$. The pseudo-code of
this is shown in Algorithm \ref{alg2}.

\begin{algorithm}
\caption{ KMC 2}
\label{alg2}
\begin{algorithmic}[0]
\State{{\bf{Input}}: $\mathbf{r}_0$, $t_0$}
\State{\bf{Output}: $\mb{r}, t$}
\State $t\gets t_0$
\State $\mathbf{r}\gets\mathbf{r}_0$
\Repeat
\State choose a suitable domain $G'$ with $\mb{r}\in G'$
\State $t_a\gets$ random number according to $\rho^D_{a}(\cdot|t)$ on $G'$
\State $t_b\gets$ random number according to 
$\rho^D_b(\cdot|\mathbf{r},t)$ on $G'$
\If{($t_a<t_b$)}
\State $\mathbf{r}\gets$ rand. position according to 
$\rho^D_n(\cdot|t_a,\mb{r},t)$ on $G'$
\Else
\State $\mathbf{r}\gets$ random position at the boundary $\partial G'$
\State $\quad\quad\,$ according to $\rho^D_f(\cdot|t_b,\mb{r},t)$
\EndIf
\State $t\gets \text{min}(t_a,t_b)$
\Until{$(\,\mb{r}\in \partial G \,\, \text{or}\,\, t_a<t_b\, )$}
\State\Return $(\mb{r},t)$
\end{algorithmic}
\end{algorithm}

\begin{figure}
\includegraphics[width=0.48\textwidth]{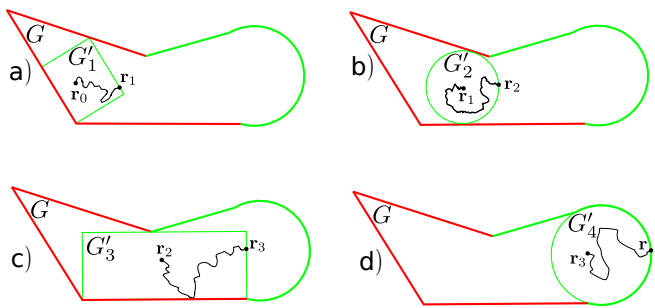}
\caption{Illustration of the method: Absorbing boundaries are shown in 
green, reflecting ones in red. In all situations a)-d) the case 
$t_a>t_b$ is sketched, otherwise the algorithm would stop earlier. 
Depending on the position of the particle the shape of the chosen domain 
$G'$ varies (rectangles and circles).}
\label{pro_box_choice}
\end{figure}

A sketch of the method in a case where the particle is absorbed at the 
boundary is shown in
Fig. \ref{pro_box_choice}.

For a given domain $G$, the efficiency of the method depends on the
choices of $G'$. Ideally, one chooses $G'$ from the list of
possibilities in a way that maximizes the expectation value of
$t_b$. However, it also takes more time to look for this special subset
and eventually calculate the random numbers for this situation. In
the cases of a particle in the middle of a circle (sphere) or in the
middle of a square (cube) the random numbers can be generated very
fast. Hence, in some situations it might be better to use smaller
domains $G'$ than in principle possible.

Up to this point, there is no approximation involved,
but depending on the shape of $\partial G$ a problem occurs: If the
particle approaches an absorbing part of the boundary of $G$, it will 
always automatically approach an absorbing part of the boundary
of the chosen $G'$, too, as $G'\subset G$. In consequence, the expectation
value of $t_b$ will decrease, the stopping condition $t_a< t_b$
becomes more and more unlikely and the time incrementations in $t$
will become smaller and smaller, if the condition $\mb{r}\in \partial
G$ is not fulfilled. But $\mb{r}\in \partial G$ can only be true, if
the intersection of $\partial G$ and $\partial G'$ contains more than
just single points. The same problem occurs for reflecting boundaries
of $G$. As the choice of $G'$ is limited, we sometimes have to
approximate $\partial G$ by a polygon in order to avoid a critical
slowing down of the algorithm. However, it is important to mention
that we can always choose the accuracy of the approximation by the
choice of the polygon.

\subsection{Maximum distance method}

Depending on the shape of $\partial G$ close to the position of the 
particle, there is sometimes a better way of propagating the particle
than it is shown in the subsection above. \cite{tenwolde1,tenwolde2}
introduced this idea for the particle's short time behavior in the
context of particle-particle interaction, but it can be modified for a
usage in our context. It is based on the assumption that there is a
maximum distance $\Delta r$, which the particle does not reach within
a time $\Delta t$. Hence, within this time interval $\Delta t$, only
the intersection of $G$ with a neighborhood of radius $\Delta r$
matters. Of course, this assumption is an approximation since there is
a non-vanishing probability that the particle leaves this neighborhood
within $\Delta t$. However, it is possible to control the accuracy by
the definition of $\Delta r$ via a parameter $\gamma$. We define:

\begin{eqnarray}
\Delta r=\gamma \sqrt{D\,\,\Delta t}\;.
\end{eqnarray}

As $\gamma$ increases, it is more and more unlikely for the particle to
violate the assumption. More precisely it is even possible to give an
upper boundary for failing the assumption by studying the first
passage-process to the boundary of a particle that starts in the
middle of a circle (2d) or a sphere (3d) with radius $\Delta r$ and
calculating the probability $w(\gamma)$ for not having reached the
boundary within $\Delta t$.

In 2 dimensions we obtain:
\begin{eqnarray}w_{2d}(\gamma)=2\sum_{n=1}^{\infty} \frac{1}{\alpha_n}
\frac{1}{J_1(\alpha_n)} e^{-\frac{\alpha_n^2}{\gamma^2}}, 
\label{assumption_correct_2d}
\end{eqnarray}

where $\alpha_n$, $n\in\mathbbm{N}$ are the roots of the Bessel
function $J_0$ (see appendix and be aware of the slightly different
notation). In the following tabular the corresponding values are
calculated for some $\gamma$.\\

\begin{tabular}{|r||r|r|r|r|r|r|r|r|r|} \hline
$\gamma$ & 2 & 3 &4 & 6 & 7& 9\\\hline
$1- w_{2d}(\gamma)$ & 0.623 & 0.193 & 0.0347& 2.41e-4 & 9.39e-06 & 
3.90e-9 \\\hline
\end{tabular}\vspace*{0.3cm}

In 3 dimensions we get (see appendix):
\begin{eqnarray}
w_{3d}(\gamma)=2\sum_{n=1}^{\infty} (-1)^{n+1} 
e^{-\left(\frac{n\pi}{\gamma}\right)^2}\;.
\label{assumption_correct_3d}
\end{eqnarray}

In the following tabular the corresponding values are calculated for 
some $\gamma$.\\
\begin{tabular}{|r||r|r|r|r|r|r|r|r|r|} \hline
$\gamma$ & 2 & 3 &4 & 6 & 7& 9\\\hline
$1- w_{3d}(\gamma)$ & 0.830 & 0.357 & 0.0827& 8.36e-4 & 3.78e-05 & 
1.63e-8 \\\hline
\end{tabular}\vspace*{0.3cm}

Consequently, for a choice of $\gamma$ in the range of $7-9$
one is on the safe side for all practical purposes, where also
other numerical error sources (quality of the random number generator,
rounding errors) come into play.

\begin{figure}
\begin{center}
\includegraphics[width=0.4\textwidth]{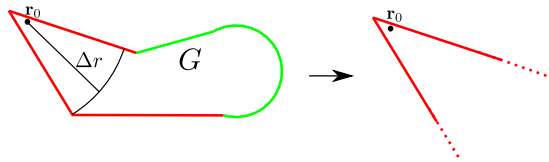}
\caption{For $t<{\Delta r^2}/{\gamma^2\,D}$, the particle is assumed
not to cross the black circle line. So the solution of an infinite 
sector with
reflecting boundaries can be used.}
\label{max_dist_bild}
\end{center}
\end{figure}

This gives the possibility to use analytic solutions of
Eq. (\ref{Dglnurdiff}) of domains which coincide with $G$ only in the
neighborhood of $\Delta r$. The example of Fig. \ref{max_dist_bild}
shows the left part of the domain from Fig. 
\ref{pro_box_choice}. Choosing $\Delta r$ in the shown way, the
analytically known solution of an infinite sector (see appendix) can
be used, as long as $t<{\Delta r^2}/{\gamma^2\,D}$.

Hence, if the particle is neither annihilated nor absorbed within
$\Delta t$, the particle will stay diffusive and a random pair
$(\mb{r},\Delta t)$ must be created for the particle's new
position. In order to avoid repetitions, we skip the pseudo-code details
here, as they will be shown in the next section in a more general
case.\vspace*{0.5cm}

In most situations it is much better to use the subset method as its
time-increments are generally much larger. But in a situation
like the one sketched in Fig. \ref{max_dist_bild}, the particle is very
close to the reflecting boundaries and no suitable large domain $G'$
is available. In consequence, $G'$ would be very small, leading to 
a very small $t_b$ on average.

\section{Inhomogeneous annihilation rate}
\label{inhoratsec}

The last section showed how to find a solution for an arbitrary domain
$G$ by solving the problem in several steps in smaller domains
$G'$. Hence, without loss of generality, we now assume the ability to
sample random numbers according to $\rho_b^D$, $\rho_f^D$ and
$\rho_n^D$ directly.

If the annihilation rate becomes inhomogeneous,
Eq. (\ref{sol_t_dep_rate}) is not a solution of (\ref{Dglallgemein})
anymore. The annihilation-time is now correlated to the complete path
of the particle, thus the method presented in the previous section
will not work. In this section we present a way to
overcome this problem for arbitrary rates $k(\mb{r},t)$ without any 
additional approximations. The following method starts with the
pair ($\mb{r}_0,t_0$) and generates a series of random pairs
($\mb{r}_i,t_i$). The last pair of this series will become the new
$(\mb{r},t)$.

For all $t>t_0$ we define the spatially homogeneous but
time dependent upper bound for the annihilation rates

\begin{equation}
k_m(t)= {\rm max}_{\mb{r}\in G}\{k(\mb{r},t)\}\; .
\end{equation}

The density $\rho_a^D$ with the rate $k_m(t)$ is denoted by $\rho_m$
in the following:

\begin{equation}
\rho_m(t|t_0)=-\frac{d}{dt} \left[ e^{-\int_{t_0}^t 
k_m(t')dt'}\right]\; .
\end{equation}

We sample a candidate pair $(\mb{r}_1,t_1)$ as shown in
Algorithm \ref{alg1}. For $\mb{r}_1\in \partial G$ the particle is absorbed at
the boundary, i.e. the first candidate is accepted. Otherwise we
compare the ratio $k(\mb{r}_1,t_1)/k_m(t_1)$ to a uniformly
distributed random number $x$ in $[0,1]$. If
$k(\mb{r}_1,t_1)/k_m(t_1)\geq x$, the particle is
annihilated, i.e. the first candidate is also accepted, else we
generate a new candidate pair $(\mb{r}_2,t_2)$ under the condition of
having been at position $\mb{r}_1$ at time $t_1$. This can be
continued until the particle is absorbed at the boundary of $G$ or
annihilated.

The algorithm can also be used to sample the complete probability
density $P(\mb{r},t|\mb{r}_0,t_0)$: If no candidate is accepted until
an arbitrarily chosen time $\tm=t$ is reached, the algorithm returns a
random position of the still diffusive particle, i.e a pair
$(\mb{r},\tm)$ whose statistics is given by
$P(\mb{r},\tm|\mb{r}_0,t_0)$. Also in case one wants to use the
maximum distance method, the time $\tm$ has to be chosen
appropriately. If a break at $\tm$ is not wanted, one simply sets
$\tm=\infty$. A pseudo-code description is shown in Algorithm
\ref{alg3}.

\begin{algorithm}
\caption{ KMC 3}
\label{alg3}
\begin{algorithmic}[0]
\State{{\bf{Input}}: $\mathbf{r}_0$, $t_0,\,t_{\text{max}}$, $k_m(t)$}
\State{\bf{Output}: $\mathbf{r}, t$}
\State $t\gets t_0$
\State $\mathbf{r}\gets\mathbf{r}_0$
\Repeat
\State $t_a\gets$ random number according to $\rho_{m}(\cdot|t)$
\State $t_b\gets$ random number according to $\rho^D_b(\cdot|\mathbf{r},t)$
\If{$(t_{\text{max}}<\text{min}(t_a,t_b))$}
\State $\mathbf{r}\gets$ random position according to 
$\rho^D_n(\cdot|\tm,\mb{r},t)$
\State $t\gets t_{\text{max}}$
\Else
\If{($t_a<t_b$)}
\State $\mathbf{r}\gets$ random position according to 
$\rho^D_n(\cdot|t_a,\mb{r},t)$
\Else
\State $\mathbf{r}\gets$ random position at the boundary $\partial G$
\State $\quad\quad\,$ according to $\rho^D_f(\cdot|t_b,\mb{r},t)$
\EndIf
\State $t\gets \text{min}(t_a,t_b)$
\EndIf
\Until{$\left((\frac{k(\mathbf{r},t)}{k_m(t)}\geq\text{ran}[0,1]){\,\,\rm 
or\,\,}(t_a>t_b) {\,\,\rm or\,\,} (t=t_\text{max})\right)$}
\State\Return $(\mathbf{r},t)$
\end{algorithmic}
\end{algorithm}

\subsection{Proof of correctness of Algorithm \ref{alg3}}

The basic mechanism by which the algorithm handles a spatially varying
annihilation rate $k(\mb{r})$ is to generate trial annihilation
positions using the propagator for a spatially constant (but maximal)
annihilation rate $k_m$. The annihilation is then accepted with the
local probability $k(\mb{r})/k_m$ for. At first sight it appears
counter-intuitive that this local procedure actually gives the correct
statistics, since the probability to propagate a particle from
$\mb{r_0}$ to $\mb{r}$ depends on the complete annihilation rate
landscape in between and around. Why is it sufficient to probe
$k(\mb{r})$ locally at one or a few positions generated by the
algorithm?

Before we answer this question rigorously by proving that it is indeed
sufficient, we give an intuitive argument why one might expect the
procedure to be correct: The stronger the spatial variation of
$k(\mb{r})$ is in $G$ the larger the maximum rate $k_m$ will be. A
large constant annihilation rate $k_m$ gives rise to a particle
propagator that forbids large hops, which implies that the algorithm
will produce many small intermediate hops and after each hop evaluates
$k(\mb{r})$. In this way the algorithm explores stochastically the
annihilation landscape on finer or coarser length scales depending on
the variation of $k(\mb{r})$. If for instance $k(\mb{r})=0$
everywhere in $G$ except in a small restricted region, where it is
$k(\mb{r})=k_1>0$, thus $k_m=k_1$. Then the algorithm will explore
the complete region $G$ with a hop size that is characteristic for
the restricted region with the non-vanishing annihilation rate. In the
end this yields the correct statistics for the whole region, which
we will prove now.

We will prove that the statistic of the output pairs
($\mb{r},t$) satisfy the probability flows $f_a$ and $f_b$ for
$t<\tm$. Then the case $t=\tm$ (particle is still diffusive at
time $\tm$) occurs with the correct probability, too. We also prove that the
statistic of output pairs $(\mb{r},\tm)$ coincides with
$P(\mb{r},\tm|\mb{r}_0,t_0)$.

The algorithm will stop after a (unknown) number $i+1$ ($i\in\mathbbm{N}$)
of loop-runs (see Algorithm \ref{alg3}). The probability density for
being annihilated after $i+1$ loop-runs at time $t$ at the position
$\mb{r}$ is denoted by $A_i(\mb{r},t|\mb{r}_0,t_0)$. Analogously the
probability density for being absorbed at the boundary after $i+1$
loop-runs at time $t$ at the position $\mb{r}$ is denoted by
$B_i(\mb{r},t|\mb{r}_0,t_0)$. The probability density for stopping
after i+1 loop-runs, still being in the diffusive state at $\tm$ and
being located at $\mb{r}$ is denoted by
$W_i(\mb{r},\tm|\mb{r}_0,t_0)$. As the number of loop-runs is a
disjoint decomposition, we can sum over $i$ to obtain the total
densities for the corresponding events:

\begin{eqnarray}
A(\mb{r},t|\mb{r}_0,t_0)&=&\sum_{i=0}^\infty 
A_i(\mb{r},t|\mb{r}_0,t_0)\; ,
\end{eqnarray}
\vspace*{-0.4cm}
\begin{eqnarray}
B(\mb{r},t|\mb{r}_0,t_0)&=&\sum_{i=0}^\infty 
B_i(\mb{r},t|\mb{r}_0,t_0)\; ,
\end{eqnarray}
\vspace*{-0.4cm}
\begin{eqnarray}
W(\mb{r},\tm|\mb{r}_0,t_0)&=&\sum_{i=0}^\infty 
W_i(\mb{r},\tm|\mb{r}_0,t_0)\; .
\end{eqnarray}

Starting with $i=0$, we compute $W_i$, $A_i$, $B_i$:
\begin{itemize}[align=left,labelwidth=\widthof{i},leftmargin=\labelwidth+\labelsep] 

\item[$i=0$:]$\,$\\
$A_0$: For this event $t_a$ has to be smaller than $t_b$, which delivers 
the second factor in the following product. The third factor 
belongs to the choice of the position and the last one arises from the 
exit-condition of the algorithm's loop:
\begin{eqnarray}
A_0(\mb{r},t|\mb{r}_0,t_0)=\rho_m(t|t_0) \left(\inuov{t}{\infty}{t_b} 
\rho^D_b(t_b|\mb{r}_0,t_0) 
\right)\nnn\cdot\internxtxn{\mb{r}}{t}{\mb{r}_0,t_0} \cdot 
\frac{k(\mb{r},t)}{k_m(t)}\nnn
=k(\mb{r},t) e^{-\int_{t_0}^{t}k_m(t')dt'} P_D(\mb{r}, 
t|\mb{r}_0,t_0)\; .
\end{eqnarray}
$B_0$: An analogous procedure delivers
\begin{eqnarray}
B_0(\mb{r},t|\mb{r}_0,t_0)=\rho^D_b(t|\mb{r}_0,t_0) 
e^{-\int_{t_0}^{t}k_m(t')dt'}&\nnn 
\cdot\randxtxn{\mb{r}}{t}{\mb{r}_0,t_0}\nonumber&\nnn
=-D e^{-\int_{t_0}^{t}k_m(t')dt'} \nabla P_D\vrtubrntn 
\cdot\mb{n}_\mb{r} &.
\end{eqnarray}

$W_0$: If the particle reaches the time $\tm$ in the first loop-run, 
$t_a$ and $t_b$ have to be larger than $\tm$. Thus $W_0$ is the product 
of these two independent probabilities with the spatial density 
$\rho^D_n(\mb{r},\tm|\mb{r}_0,t_0)$:

\begin{eqnarray}
W_0(\mb{r},\tm|\mb{r}_0,t_0)=\left(\inuov{\tm}{\infty}{t_a} 
\rho_m(t_a|t_0)\right)\quad\quad\quad\,\,\nnn \cdot 
\left(\inuov{\tm}{\infty}{t_b} 
\rho^D_b(t_b|\mb{r}_0,t_0)\right)\cdot\internxtxn{\mb{r}}{\tm}{\mb{r}_0,t_0}\nnn
=e^{-\int_{t_0}^{\tm}k_m(t')dt'} P_D(\mb{r},\tm|\mb{r}_0,t_0)\;.\quad
\end{eqnarray}

\item[$i=1$:]$\,$\\ As the algorithm will pass the loop twice here, we 
have to sum/integrate over all weighted pairs $(\mb{r_1},t_{1})$, which 
will be achieved in the first loop-run. Since the algorithm will only 
continue with a new loop if $t_a$ is smaller than $t_b$, for the first 
loop the factors and integrals look the same for all cases. The factors 
of the final loop can be taken from the individual factors of $i=0$ with 
the starting position $\mb{r}_1$ and the time $t_1$ instead of $\mb{r}_0$ and 
$t_0$. Defining the probability that the algorithm denies a candidate 
pair $(\mb{r},t)$
\begin{eqnarray}
q(\mb{r},t)=\ablfac{\mb{r}}{t}\; ,
\end{eqnarray}
we get:

\end{itemize}
\begin{widetext}
\begin{eqnarray}
A_1(\mb{r},t|\mb{r}_0,t_0)&=&\inuov{t_0}{t}{t_{a1}} 
\rho_m(t_{a1}|t_0)\inuov{t_{a1}}{\infty}{t_{b1}} 
\rho_b^D(t_{b1}|\mb{r}_0,t_0)
\int_G d\mb{r}_1 
\internxtxn{\mb{r}_1}{t_{a1}}{\mb{r}_0,t_0}\,q(\mb{r}_1,t_{a1})\,A_0(\mb{r},t|\mb{r}_1,t_{a1}) 
\nnn
&=&k(\mb{r},t)\,e^{-\int_{t_0}^{t}k_m(t')dt'}\inuov{t_0}{t}{t_{a1}}\int_G d\mb{r}_1\,P_D(\mb{r}_1,t_{a1}|\mb{r}_0,t_0) 
k_m(t_{a1})\,q(\mb{r}_1,t_{a1})\,P_D(\mb{r},t|\mb{r}_1,t_{a1})\quad
\end{eqnarray}
\vspace*{-0.2cm}
\begin{eqnarray}
B_1(\mb{r},t|\mb{r}_0,t_0)&=&\inuov{t_0}{t}{t_{a1}} 
\rho_m(t_{a1}|t_0)\inuov{t_{a1}}{\infty}{t_{b1}} 
\rho_b^D(t_{b1}|\mb{r}_0,t_0)
\int_G d\mb{r}_1 
\internxtxn{\mb{r}_1}{t_{a1}}{\mb{r}_0,t_0}\,q(\mb{r}_1,t_{a1})\, 
B_0(\mb{r},t|\mb{r}_1,t_{a1})\nnn
&=&-D\,\,e^{-\int_{t_0}^{t}k_m(t')dt'} \inuov{t_0}{t}{t_{a1}}\int_G 
d\mb{r}_1\,P_D(\mb{r}_1,t_{a1}|\mb{r}_0,t_0) 
k_m(t_{a1})\,q(\mb{r}_1,t_{a1})\,\nabla 
P_D(\mb{r},t|\mb{r}_1,t_{a1})\cdot\mb{n}_\mb{r}
\end{eqnarray}
\vspace*{-0.2cm}
\begin{eqnarray}
&&\hspace*{-1cm}W_1(\mb{r},\tm|\mb{r}_0,t_0)\\&=&\inuov{t_0}{\tm}{t_{a1}} \rho_m(t_{a1}|t_0)\inuov{t_{a1}}{\infty}{t_{b1}} 
\rho_b^D(t_{b1}|\mb{r}_0,t_0)
\int_G d\mb{r}_1 
\internxtxn{\mb{r}_1}{t_{a1}}{\mb{r}_0,t_0}\,q(\mb{r}_1,t_{a1})\,W_0(\mb{r},\tm|\mb{r}_1,t_{a1})\nnn
&=&e^{-\int_{t_0}^{\tm}k_m(t')dt'}\inuov{t_0}{\tm}{t_{a1}}\int_G 
d\mb{r}_1\,P_D(\mb{r}_1,t_{a1}|\mb{r}_0,t_0) 
k_m(t_{a1})\,q(\mb{r}_1,t_{a1})\,P_D(\mb{r},\tm|\mb{r}_1,t_{a1})
\end{eqnarray}
For $i\geq 1$ we introduce the definitions
\begin{eqnarray}
h_i(\mb{r},t|\mb{r}_0,t_0)=\underset{t_0\leq t_1\leq ...\leq t_i \leq 
t}{\int \left( \prod_{j=1}^i dt_j \right)}\,\, k_m(t_j) \,\, \int_{G^n} 
\left(\prod_{k=1}^i d\mb{r}_k\right)\,\,Q_i\left(\left\lbrace 
\left(\mb{r}_l,t_l\right)\right\rbrace_{l=0..i}\right)P_D(\mb{r},t|\mb{r}_i,t_i)\; 
,\label{def_hi}\\
\text{with}\quad Q_i\left(\left\lbrace 
\left(\mb{r}_l,t_l\right)\right\rbrace_{l=0..i}\right)=\prod_{l=1}^i 
P_D(\mb{r}_l,t_l|\mb{r}_{l-1},t_{l-1}) \cdot q(\mb{r}_l,t_l)\; .
\end{eqnarray}
Finally, defining
\begin{equation}
h_0\vrtubrntn = P_D\vrtubrntn \; ,
\end{equation}
one inductively gets for $i\geq 0$:
\begin{eqnarray}
A_i(\mb{r},t|\mb{r}_0,t_0)&=&k(\mb{r},t) e^{-\int_{t_0}^{t}k_m(t')dt'} 
h_i(\mb{r},t|\mb{r}_0,t_0)\; , \\
B_i(\mb{r},t|\mb{r}_0,t_0)&=&-D\,e^{-\int_{t_0}^{t}k_m(t')dt'}\,\nabla 
h_i(\mb{r},t|\mb{r}_0,t_0)\cdot\mb{n}_\mb{r}\; , \\
W_i(\mb{r},\tm|\mb{r}_0,t_0)&=&e^{-\int_{t_0}^{\tm}k_m(t')dt'} 
h_i(\mb{r},\tm|\mb{r}_0,t_0)\; .
\end{eqnarray}
Hence, the total probability densities can be written as
\begin{eqnarray}
A(\mb{r},t|\mb{r}_0,t_0)&=&k(\mb{r},t)\,\, \tilde{P}\vrtubrntn\; , \\
B(\mb{r},t|\mb{r}_0,t_0)&=&-D\,\, \nabla 
\tilde{P}\vrtubrntn\cdot\mb{n}_\mb{r}\; , \\
W(\mb{r},\tm|\mb{r}_0,t_0)&=& \tilde{P}(\mb{r},\tm|\mb{r}_0,t_0)\ ,
\end{eqnarray}
with
\begin{eqnarray}
\tilde{P}\vrtubrntn=e^{-\int_{t_0}^{t}k_m(t')dt'} \sum_{i=0}^\infty 
h_i(\mb{r},t|\mb{r}_0,t_0)\; .
\end{eqnarray}
Comparing this with the definitions of $f_a$ and $f_b$, it remains to 
show that $\tilde{P}\vrtubrntn\stackrel{!}{=} P\vrtubrntn$, i.e. 
$\tilde{P}\vrtubrntn$ has to satisfy
Eq. (\ref{Dglallgemein}) with the initial condition 
$\tilde{P}(\mb{r},t_0|\mb{r}_0,t_0)=\delta (\mb{r}-\mb{r}_0)$. As all 
$h_i$ with $i\geq 1$ vanish for $t=t_0$, the initial condition is simply 
fulfilled by the definition of $h_0$. For the time-derivative of $h_i$ 
one inductively gets for $i\geq 1$:

\begin{eqnarray}
\dot{h}_i(\mb{r},t|\mb{r}_0,t_0)&=&k_m(t)\,q(\mb{r},t)\cdot 
h_{i-1}\vrtubrntn\nnn &&+\underset{t_0\leq t_1\leq ...\leq t_i \leq 
t}{\int \left(\prod_{j=1}^i dt_j\right)}\,\, k_m(t_j) \int_{G^n} 
\left(\prod_{k=1}^i d\mb{r}_k\right) Q_i\left(\left\lbrace 
\left(\mb{r}_l,t_l\right)\right\rbrace_{l=0..i}\right)\dot{P}_D(\mb{r},t|\mb{r}_i,t_i) 
\label{abl_hi}
\end{eqnarray}
Hence, the time-derivative of $\tilde{P}\vrtubrntn$ satisfies
\begin{eqnarray}
\pdou{{\tilde{P}}(\mb{r},t|\mb{r}_0,t_0)}{t}&=&-k_m(t)\,\tilde{P}\vrtubrntn+ 
e^{-\int_{t_0}^{t}k_m(t')dt'} \left\lbrace \dot{P}_D\vrtubrntn + 
\sum_{i=1}^\infty \dot{h}_i(\mb{r},t|\mb{r}_0,t_0) \right\rbrace \\
&=&-k_m(t)\,\tilde{P}\vrtubrntn + e^{-\int_{t_0}^{t}k_m(t')dt'} 
\left\lbrace \dot{P}_D\vrtubrntn + k_m(t)\,q(\mb{r},t) \sum_{i=1}^\infty 
h_{i-1}\vrtubrntn \right. \nnn
&& \left. + \sum_{i=1}^\infty \underset{t_0\leq t_1\leq ...\leq t_i \leq 
t}{\int \left(\prod_{j=1}^i dt_j\right)}\,\, k_m(t_j) \,\, 
\int_{G^n}\left( \prod_{k=1}^i d\mb{r}_k 
\right)\,\,Q_i\left(\left\lbrace 
\left(\mb{r}_l,t_l\right)\right\rbrace_{l=0..i}\right)\dot{P}_D(\mb{r},t|\mb{r}_i,t_i) 
\right\rbrace \\
&=&-k(r,t)\tilde{P}\vrtubrntn + e^{-\int_{t_0}^{t}k_m(t')dt'} 
\left\lbrace \dot{P}_D\vrtubrntn + \right. \nnn
&& \left. \sum_{i=1}^\infty \underset{t_0\leq t_1\leq ...\leq t_i \leq 
t}{\int \left(\prod_{j=1}^i dt_j\right)}\,\, k_m(t_j) \,\, 
\int_{G^n}\left( \prod_{k=1}^i d\mb{r}_k\right) 
\,\,Q_i\left(\left\lbrace 
\left(\mb{r}_l,t_l\right)\right\rbrace_{l=0..i}\right)\dot{P}_D(\mb{r},t|\mb{r}_i,t_i) 
\right\rbrace\; .
\end{eqnarray}
Thus, using Eq. (\ref{Dglnurdiff}), it follows:
\begin{eqnarray}
\pdou{{\tilde{P}}(\mb{r},t|\mb{r}_0,t_0)}{t}&=&-k(\mb{r},t)\tilde{P}\vrtubrntn 
+ e^{-\int_{t_0}^{\tm}k_m(t')dt'} D\,\,\Delta \left\lbrace 
\sum_{i=0}^\infty h_i\vrtubrntn \right\rbrace \nnn
&=& -k(\mb{r},t)\tilde{P}\vrtubrntn + D\,\, \Delta 
\tilde{P}(\mb{r},t|\mb{r}_0,t_0)\; ,
\end{eqnarray}
which is exactly Eq. (\ref{Dglallgemein}). Hence, the correctness of the 
Algorithm \ref{alg3} is proven.\vspace*{0.6cm}
\end{widetext}

\section{Example}
\label{simuerg}

This section presents a two-dimensional application example of the
algorithm. It is designed to demonstrate how the algorithm handles a
situation in which its correctness is most counter-intuitive: We choose
the annihilation rate to be non-vanishing just inside a restricted
region, a circle, where it oscillates in time and varies spatially.
For a chosen test-setup, we compare its results with the solution of a
commercial FEM (finite element method) routine.

\begin{figure}
\begin{center}
\includegraphics[width=0.45\textwidth]{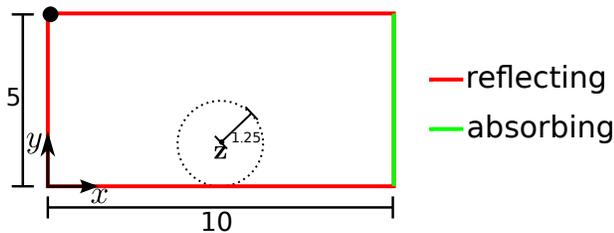}
\caption{sketch of the simulation setup: The particle starts its
diffusive motion at the upper left corner. It can either be absorbed
at the right wall or annihilated within the drawn circle.}
\label{2dgeo_example}
\end{center}
\end{figure}

At $t_0=0$ the diffusing particle ($D=1$) is located at the position
$\mb{r}_0=(0;5)$ within a rectangle of size $10\times 5$. The right
boundary is chosen to be absorbing, all other boundaries are
reflecting. A strongly anisotropic time dependent annihilation rate $
k(\mb{r},t)$ is chosen to be

\begin{eqnarray}
k(\mb{r},t)=\left\lbrace{\small \begin{matrix} 
3|\cos^3\left(\frac{t}{2}\right)|\cdot\left(c^2-||\mb{r}-\mb{z}||^2\right),& 
||\mb{r}-\mb{z}|| <c \\ 0&\hspace*{-0.3cm},\,\,\, ||\mb{r}-\mb{z}|| \geq 
c \end{matrix}} \right.\; ,\,\,\,
\label{annirate_exmple}
\end{eqnarray}
with $\mb{z}=(5;1.25)$ and $c=1.25$. Fig. \ref{2dgeo_example} presents 
a sketch of the described setup.

On the one hand the problem has been solved numerically by applying a
commercial FEM solver with a very fine triangulation ($>600000$
elements) to Eq. (\ref{Dglallgemein}). In the following this solution
is denoted by $P_F(\mb{r},t)$. On the other hand the Monte-Carlo
algorithm has been applied to the problem in $4.2 \times 10^8$
samples. In principle it is not necessary to use the subset method
here, as the analytic solution of Eq. (\ref{Dglnurdiff}) is known for
the rectangle (see appendix), from which all occurring densities can
be sampled. Nevertheless it has been used, as it increases the speed
of the algorithm dramatically: For all highly anisotropic
annihilation rates the ratio $k/k_m$ in Algorithm \ref{alg3} will
mostly be very small (in our case even 0). Hence, a lot of loop runs
with just small time incrementations will on average be needed for an
event. Restricting the movement of the particle temporally to a subset of
$G$ (subset method) ensures the possibility of choosing a smaller
$k_m(t)$, which reduces the number of loop-runs in Algorithm \ref{alg3}
dramatically. Using this in our case it takes around 40 minutes for
$10^6$ samples on a single core with $3.4$ GHz.

\begin{figure}[H]
\begin{center}
\includegraphics[width=0.45\textwidth]{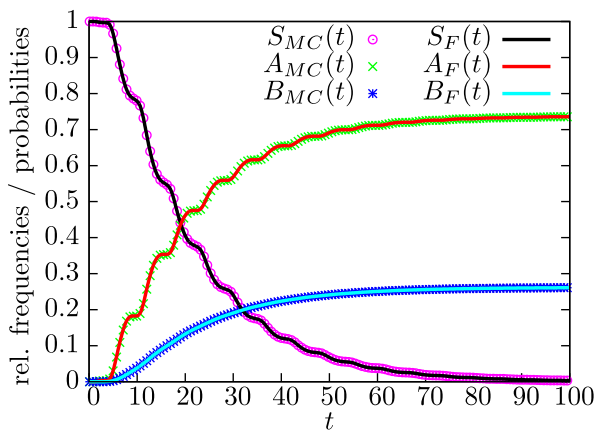}
\caption{Comparison of the relative frequencies derived from Algorithm 
\ref{alg3} and the numerically integrated probabilities for the events of 
still being diffusive ($S$), having already been annihilated ($A$) and
having already been absorbed at the right wall ($B$).}
\label{kummugraph}
\end{center}
\end{figure}

Firstly, the relative frequencies for the times of an event and the
kind of the event were counted. $S_{MC}(t)$ denotes the relative
frequency of having had no event until time $t$. It has to be compared
with the value of $S_{F}(t)=1-\int_0^t \rho_e (t'|\mb{r}_0,t_0) dt'$,
which was derived  numerically from the FEM solution. $A_{MC}(t)$
denotes the relative frequency of having had an annihilation event
before time $t$. It is compared to $A_{F}(t)=\int_0^t
\alpha(t'|\mb{r}_0,t_0) dt'$. $B_{MC}(t)$ denotes the relative
frequency of being absorbed at the right boundary before time $t$. It
is compared to $B_{F}(t)=\int_0^t \beta(t'|\mb{r}_0,t_0) dt'$. A plot
of these quantities is shown in Fig. \ref{kummugraph}.

It shows an almost perfect coincidence of all corresponding
quantities. The maximum relative deviation is about 1$\%$ in all
curves. Keeping in mind that $S_{F}(t)$, $A_{F}(t)$ and $B_{F}(t)$ are
calculated by a numerical time-integration of a numerical spatial
integration of a numerical solution of Eq. (\ref{Dglallgemein}), these
small deviations are explainable. More precisely,
$S_{F}(t)+A_{F}(t)+B_{F}(t)=1$ has to hold for all times, but the
numerical discrepancy in this sum is also about 1$\%$ at maximum.

\begin{figure*}
\includegraphics{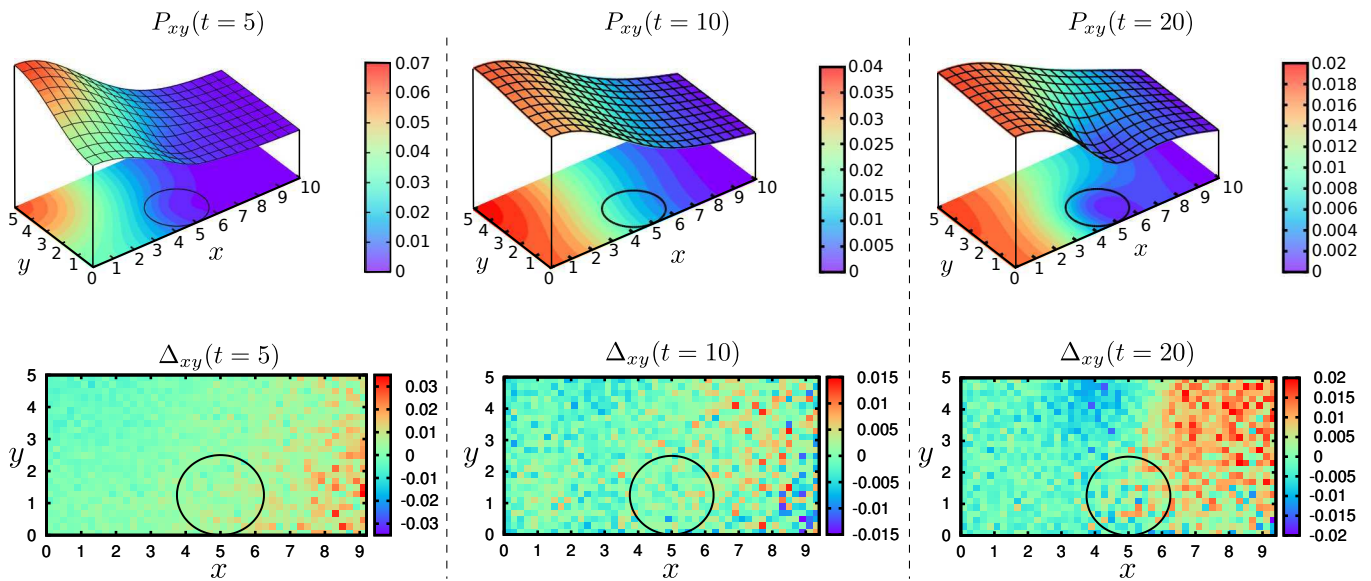} 
\caption{{\bf Top}:
The probability density $P_{xy}(t)=P(\mb{r},t|\mb{r}_0,t_0)$
for the two-dimensional geometry depicted in Fig.\ \ref{2dgeo_example}
and the annihilation rate given by Eq.\ (\ref{annirate_exmple})
generated by the intermediate positions of Algorithm \ref{alg3}
for the times $t_1=5$, $t_2=10$ and $t_3=20$. The position of
the annihilation zone with oscillating strength is indicated by
the full circle. {\bf Bottom}: Relative difference $\Delta_{xy}(t)$
between the Monte Carlo result $P_{xy}(t)$ and the numerical
solution of the corresponding annihilation-diffusion
Eq. (\ref{Dglallgemein}) for the times $t_1=5$ , $t_2=10$
and $t_3=20$.}
\label{example_graphs}
\end{figure*}

Secondly, we want to compare the spatial distribution of the
particle's position from the KMC algorithm to $P_F(\mb{r},t)$ for
three characteristic times: $t_1=5$, $t_2=10$, $t_3=20$. Hence, the
rectangle is divided in $2N \times N$ squares $s_{xy}$ ($x=i\cdot
\frac{5}{N}$, $y=j\cdot \frac{5}{N},\,\, i=0,\, 1, \ldots
2N-1,\,\,j=0,\, 1, \ldots N-1$) and the relative frequency $h_{xy}(t)$
for being at the square $s_{xy}$ is counted for $t_1$, $t_2$ and
$t_3$. Technically this has been done by setting $t_{max}=t_i$ $(i\in
\left\lbrace 1,2,3\right\rbrace)$ in Algorithm \ref{alg3}. The quotient of
$h_{xy}(t)$ and the area of a square element is denoted by
$P_{xy}(t)$. This density converges to the solution of
Eq. (\ref{Dglallgemein}) in the limits of increasing sample numbers
and $N\rightarrow \infty$. The upper panel in Fig. 
\ref{example_graphs} shows the density $P_{xy}(t)$ for the chosen
times in a 3d plot for $N=50$. In order to illustrate the influence of
the annihilation within the circle, the projection on the bottom shows
isolines by discretising the density into intervals.

\begin{itemize}[align=left,labelwidth=\widthof{i},leftmargin=\labelwidth+\labelsep]
\item $t_1=5$: The probability density of the particle is still centered 
around the starting position in the upper left corner. Nevertheless its 
shape is already
influenced by the annihilation within the circle.
\item $t_2=10$: At time $t_2$ there has been almost no annihilation for 
a short period (slope of the red line in Fig. \ref{kummugraph}). 
Hence, diffusion almost equilibrated the
density gradient in y-direction, generated by the annihilation within 
the time-interval $[5;8]$.
\item $t_3=20$: At time $t_3$ relatively strong annihilation takes 
place, which even leads to a local minimum of $P_{xy}$ within the circle.
\end{itemize}
In order to quantify the local differences between the KMC result and the FEM 
result, we choose squares of size $0.2\times 0.2$ ($N=25$).
A measure for the spatially resolved relative deviation is
\begin{eqnarray}
\Delta_{xy}(t)=\frac{h_{xy}(t)-\int_{x}^{x+0.2} dx' \int_{y}^{y+0.2} dy' 
P_F(\mb{r}',t)}{\int_{x}^{x+0.2} dx' \int_{y}^{y+0.2} dy' P_F(\mb{r}',t) 
}\; .\non
\end{eqnarray}

The lower panel of Fig. \ref{example_graphs} shows $\Delta_{xy}(t)$.
For all times the deviations are small and in the range of
the numerical expectation: For $4.2\times 10^8$ samples and
$(2N)^2=10^4$ plaquettes one expects ca. $10^4$ samples per
plaquette and thus statistical fluctuation of the order of
$10^2$, i.e.\ relative fluctuation in the 1 percent range,
which is what the lower panel of Fig.\ \ref{example_graphs} confirms.

On the right side of the simulation rectangle,
where the absorbing boundary is located, the statistical error is larger for small
times, since the density is still centered around the starting 
point in the
upper left corner, giving a region $(x>8)$ with very small values of
$h_{xy}(t)$. But the aim of the Algorithm \ref{alg3} is not the
stochastic solution of Eq. (\ref{Dglallgemein}), for which
finite element methods are suitable. The aim of the algorithm is to
sample correctly \emph{next events} (annihilations or first-passages
of boundaries) according to Eq. (\ref{Dglallgemein}), which can not be
handled by a FEM routine. The example demonstrates, that this is
possible, even in cases of highly anisotropic and time dependent
annihilation rates.

\section{Discussion}

We have presented an algorithm that samples correctly the probability
distribution of a diffusing particle with a space dependent
annihilation or transformation rate $k(\mb{r})$ for arbitrary
domains. Together with first-passage time methods it can serve as the
basic building block for a kinetic Monte Carlo algorithm simulating a
general many-particle reaction-diffusion system. 

The basic idea is to generate trial moves with the exactly known
single particle Green's function for a spatially constant annihilation
rate $k_m$, which is the maximum of $k(\mb{r})$ in the current
protecting domain. With probability $k(\mb{r})/k_m$ the particle is
annihilated at the trial position ${\mb r}$, otherwise a new trial
move with initial position ${\mb r}$ is generated. The iteration
proceeds until either the particle is annihilated or the boundary of
the protecting domain is reached. In this paper we proved rigorously
the correctness of this algorithm and demonstrated its numerical
accuracy and efficiency with an illustrative example.

Important applications with a spatially varying transformation rate
include continuum models for intracellular transport (or more generally
intermittent search strategies \cite{benichou}). In intracellular
transport particles (proteins, organelles) can switch between free
diffusion and ballistic motion by molecular motor assisted movement
along cytoskeleton filaments. The density of filaments in the space
direction $\Omega$, $\rho_\Omega(\mb{r},t)$, is generally very
inhomogeneous in space and sometimes even varies over time (for
instance during cell polarization). This situation can be described by
the Fokker-Planck equation for the probability densities $P_0(\mb{r},t)$
and $P_\Omega(\mb{r},t)$ for diffusing particles and particles that
move with a constant velocity ${\bf v}_\Omega$ in direction $\Omega$,
respectively \cite{loverdo}:

\begin{eqnarray}
\frac{\partial}{\partial t} P_0(\mb{r},t)
&=&D\Delta P_0(\mb{r},t)
 -\gamma\,P_0(\mb{r},t)\int d\Omega\,\rho_\Omega(\mb{r},t)\nonumber\\ 
 && +\gamma'\int d\Omega\, P_\Omega(\mb{r},t)\label{diff}\\
\frac{\partial}{\partial t} P_\Omega(\mb{r},t) 
&=&-\nabla\cdot ({\bf v}_\Omega P_\Omega(\mb{r},t))
 +\gamma\,\rho_\Omega(\mb{r},t)P_0(\mb{r},t)\nonumber\\
 &&-\gamma'P_\Omega(\mb{r},t)\;,\label{ball}
\end{eqnarray}

where $\gamma$ and $\gamma'$ are the attachment and detachment rates
(to and from filaments), respectively. The freely diffusing particle
sees a total annihilation rate $k(\mb{r},t)=\gamma\int
d\Omega\,\rho_\Omega(\mb{r},t)$, with which it is transformed into a
ballistically moving particle with a randomly chosen direction
$\Omega$ (and velocity ${\bf v}_\Omega$) with probability
$\rho_\Omega(\mb{r},t)/\int d\Omega\,\rho_\Omega(\mb{r},t)$. The
algorithm presented in this paper handles a Monte Carlo simulation 
of the diffusion process described by (\ref{diff}), whereas the
implementation of the ballistic motion (\ref{ball}) is
straightforward.

\appendix
\section{}

This appendix presents some analytic solutions of Eq. 
(\ref{Dglnurdiff}), which have mostly been taken from \cite{carslaw}. 
Furthermore, it derives expressions for sampling according to the 
densities $\rho_b^D$, $\rho_f^D$, $\rho_n^D$ in cases where this might 
not be obvious anymore. We list only frequently used domains in one, two 
and three dimensions. 

\subsection{Particle on the interval $[0,L]$}

\begin{itemize}[align=left,labelwidth=\widthof{i},leftmargin=\labelwidth+\labelsep] 

\item absorbing on both sides:
\begin{eqnarray}
P_D(x,t|x_0,t_0)=\frac{2}{L}\sum_{n=1}^\infty 
e^{-{k_n}^2D(t-t_0)}\sin\left(k_n x\right)\sin\left(k_n x_0\right)\;, 
\nonumber
\end{eqnarray}
with $k_n=\frac{n\pi}{L}$.\\ Expressions for the probability densities 
$\rho_b^D$, $\rho_f^D$, $\rho_n^D$ and the corresponding distribution 
functions $F_b^D$, $F_f^D$, $F_n^D$ can be derived analytically.
\item reflecting on the left and absorbing on the right side:
\begin{eqnarray}
P_D(x,t|x_0,t_0)=\frac{2}{L}\sum_{n=0}^\infty 
e^{-{k_n}^2D(t-t_0)}\cos\left(k_n x\right)\cos\left(k_n x_0\right)\;, 
\nonumber
\end{eqnarray}
with $k_n=\frac{(2n+1)\pi}{2L}$.\\ Expressions for the probability 
densities $\rho_b^D$, $\rho_n^D$ and the corresponding distribution 
functions $F_b^D$, $F_n^D$ are analytically derivable. As there is only 
$x=L$ for the particle to leave the domain, it follows 
$\rho_f^D(0|t,x_0,t_0)=0$ and $\rho_f^D(L|t,x_0,t_0)=1$.
\item reflecting on both sides:
\begin{eqnarray}
P_D(x,t|x_0,t_0)=\hspace*{5cm}\nnn\frac{1}{L}\left(1+2\sum_{n=0}^\infty 
e^{-{k_n}^2D(t-t_0)}\cos\left(k_n x\right)\cos\left(k_n 
x_0\right)\right)\;, \nonumber
\end{eqnarray}
with $k_n=\frac{n\pi}{L}$.\\ Expressions for the probability density 
$\rho_n^D$ and the corresponding distribution function $F_n^D$ can be 
derived analytically. 
\end{itemize}

\subsection{Particle in a rectangle $[0,a]\times [0,b]$ and in a cuboid 
$[0,a]\times [0,b]\times [0,c]$ }

If the boundary conditions do not vary along each side, $P_D$ factorizes:
\begin{eqnarray}
P_D=P_D^a(x,t|x_0,t_0) \cdot P_D^b(y,t|y_0,t_0)\,\,\,\,\left( \cdot 
P_D^c(z,t|z_0,t_0) \right)\;, \nonumber
\end{eqnarray}
where $P_D^a$, $P_D^b$ ($P_D^c$) are given by solutions for intervals 
from the subsection above. Depending on the boundary conditions, 
$\rho_b^D$ is sampled by generating a random time for every coordinate, 
where there is at least one absorbing boundary. The smallest of these 
times has to be returned as $t_b$. The particle reaches the boundary in 
the corresponding coordinate. All other quantities are sampled as above.

\subsection{Particle in a circle of radius $R$}

\begin{itemize}[align=left,labelwidth=\widthof{i},leftmargin=\labelwidth+\labelsep] 

\item absorbing boundary:
\begin{eqnarray}
P_D(r,\varphi,t|r_0,\varphi_0,t_0)=\frac{1}{\pi R^2}\left[\sum_{n=-\infty}^\infty \cos\left( n 
\left(\varphi-\varphi_0 \right) \right)\right.\quad\,\,\nnn\cdot\left.\sum_{\alpha_n} 
e^{-\alpha_n^2\frac{D(t-t_0)}{R^2}}\frac{J_n\left( 
\alpha_n\frac{r}{R}\right) J_n\left( 
\alpha_n\frac{r_0}{R}\right)}{J_n'(\alpha_n)^2}\right]\;,\nonumber
\end{eqnarray}
where $\sum_{\alpha_n}$ denotes the infinite sum over all positive roots 
$\alpha_n$ of the Bessel function $J_n(\alpha_n)=0$.\\
The density of finding the particle at an arbitrary angle at radius $r$ 
is then given by
\begin{eqnarray}
\rho_r(r,t|r_0,t_0)&=&\int_0^{2\pi}P_D(r,\varphi,t|r_0,\varphi_0,t_0) 
r\,d\varphi\nnn &=& \frac{2}{R^2}\sum_{\alpha_0} 
e^{-\alpha_0^2\frac{D(t-t_0)}{R^2}}r\,\frac{J_0\left( 
\alpha_0\frac{r}{R}\right) J_0\left( 
\alpha_0\frac{r_0}{R}\right)}{J_1(\alpha_0)^2}\;.\nonumber
\end{eqnarray}
and the corresponding distribution function is given by
\begin{eqnarray}
F_r(r,t|r_0,t_0)&=&\int_0^r dr' \rho_r(r',t|r_0,t_0) \nnn 
&=&\frac{2}{R}\sum_{\alpha_0} 
e^{-\alpha_0^2\frac{D(t-t_0)}{R^2}}r\,\frac{J_1\left( 
\alpha_0\frac{r}{R}\right) J_0\left( 
\alpha_0\frac{r_0}{R}\right)}{\alpha_0\,J_1(\alpha_0)^2}\;.\nonumber
\end{eqnarray}
Hence, the distribution function belonging to $\rho_b^D$ is given by
\begin{eqnarray}
F_b^D(t|r_0,t_0)=1-2\sum_{\alpha_0} 
e^{-\alpha_0^2\frac{D(t-t_0)}{R^2}}\frac{J_0\left( 
\alpha_0\frac{r_0}{R}\right)}{\alpha_0\,J_1(\alpha_0)}\; .\nonumber
\end{eqnarray}
Analytic expressions for all quantities depending on $\varphi$ which are 
needed, are straightforwardly derivable by integrating the 
$\cos$-functions.\\
Having precomputed the values of $\alpha_n$, random numbers are sampled 
by inverting the occurring distribution functions numerically.\\
For a particle starting in the center of the circle the $\varphi$-
dependence becomes uniformly distributed in the interval $[0,2\pi[$ and 
$F_b^D$ simplifies to
\begin{eqnarray}
F_b^D(t|0,t_0)=1-2\sum_{\alpha_0} 
e^{-\alpha_0^2\frac{D(t-t_0)}{R^2}}\frac{1}{\alpha_0\,J_1(\alpha_0)}\; , \nonumber
\end{eqnarray}
which is used to derive Eq. (\ref{assumption_correct_2d}).
\item reflecting boundary:
\begin{eqnarray}
P_D(r,\varphi,t|r_0,\varphi_0,t_0)=\frac{1}{\pi R^2}\left[1+\sum_{n=-\infty}^\infty 
\cos\left( n \left(\varphi-\varphi_0 \right) \right)\right.\quad\nnn\cdot\left.\sum_{\alpha_n} 
e^{-\alpha_n^2\frac{D(t-t_0)}{R^2}}\frac{J_n\left( 
\alpha_n\frac{r}{R}\right) J_n\left( 
\alpha_n\frac{r_0}{R}\right)}{\left(1-\frac{n^2}{\alpha_n^2} \right) 
J_n(\alpha_n)^2}\right]\;, \non
\end{eqnarray}
where $\sum_{\alpha_n}$ denotes the infinite sum over all positive roots 
$\alpha_n$ of $J_n'(\alpha_n)=0$.\\
The density of finding the particle at an arbitrary angle at radius $r$ 
is then given by
\begin{eqnarray}
\rho_r(r,t|r_0,t_0)=\int_0^{2\pi}\rho(r,\varphi,t|r_0,\varphi_0) 
r\,d\varphi\quad\quad\quad\quad\quad\quad \nnn
=\frac{2}{R^2}\left[r+\sum_{\alpha_0} 
e^{-\alpha_0^2\frac{D(t-t_0)}{R^2}}r\,\frac{J_0\left( 
\alpha_0\frac{r}{R}\right) J_0\left( 
\alpha_0\frac{r_0}{R}\right)}{J_0(\alpha_0)^2}\right] \non
\end{eqnarray}
and the corresponding distribution function is given by
\begin{eqnarray}
F_r(r,t|r_0,t_0)=\quad\quad\quad\quad\quad\quad\quad\quad\quad\quad\quad\quad\quad\quad\nnn\frac{r^2}{R^2}+\frac{2}{R}\sum_{\alpha_0} 
e^{-\alpha_0^2\frac{D(t-t_0)}{R^2}}r\,\frac{J_1\left( 
\alpha_0\frac{r}{R}\right) J_0\left( 
\alpha_0\frac{r_0}{R}\right)}{\alpha_0 J_0(\alpha_0)^2}\; .\non
\end{eqnarray}
A distribution function for the angle $\varphi$ under the condition of 
being at radius $r$ can be derived straightforwardly by integrating the 
$\cos$-functions.
Having precomputed the values of $\alpha_n$, $r$ and $\varphi$ are 
sampled by a numerical inversion of the distribution functions. For a 
particle starting in the center of the circle the $\varphi$-dependence 
again becomes uniformly distributed in the interval $[0,2\pi[$.
\end{itemize}

\subsection{Particle in a sector of angle $\Theta$ with reflecting 
boundaries}

\begin{eqnarray}
P_D(r,\varphi,t|r_0,\varphi_0,t_0)=\dfrac{e^{-\frac{r^2+r_0^2}{4D(t-t_0)}}}{2\Theta 
D(t-t_0)}\left[I_0\left(\frac{r\,r_0}{2D(t-t_0)}\right)\right.+\nnn\left.2\sum_{n=1}^\infty 
\cos\left( n \frac{\pi\varphi}{\Theta}\right) \cos\left( n 
\frac{\pi\varphi_0}{\Theta}\right) I_\frac{n\pi}{\phi} 
\left(\frac{r\,r_0}{2Dt}\right) \right]\; ,\non
\end{eqnarray}
where $I_\omega$ denotes the modified Bessel function of order $\omega$.

The density for finding the particle at an arbitrary angle $\varphi$ at radius $r$ 
is then given by
\begin{eqnarray}
\rho_r(r,t|r_0,t_0)=\frac{r\,e^{-\frac{r^2+r_0^2}{4D(t-t_0)}}}{2 
D(t-t_0)}\left[I_0\left(\frac{r\,r_0}{2D(t-t_0)}\right)\right]\; .\non
\end{eqnarray}
As there is no analytic expression for a distribution function of
$\rho_r$ available, the usage of the inversion method would be very
slow, as the integration of $r$ would have to be done
numerically. Fortunately, $\rho_r$ does not depend on the sector angle
$\Theta$, hence the analytically known solution for $\Theta=\pi$ 
(half-plane) can be used to generate $r$ for 
all $\Theta$. A distribution function for the angle $\varphi$ under
the condition of being at radius $r$ can be derived straightforwardly
by integrating the $\cos$-function.

\subsection{Particle in a sphere of radius $R$ with absorbing boundary 
conditions }
\begin{eqnarray}
P_D(r,\varphi,\vartheta,t|r_0,\varphi_0,\vartheta_0,t_0)=\quad\quad\quad\quad\quad\quad\quad\quad\nnn
\frac{1}{2\pi R^2 \sqrt{r r_0}}\sum_{n=0}^\infty (2n+1) 
P_n(\mu(\varphi,\vartheta,\varphi_0,\vartheta_0))\nnn
\cdot\sum_{\alpha_n} e^{-\alpha_n^2\frac{D(t-t_0)}{R^2}}\dfrac{ 
J_{n+\frac{1}{2}}\left( 
\alpha_n\frac{r_0}{R}\right)J_{n+\frac{1}{2}}\left( 
\alpha_n\frac{r}{R}\right)}{\left[J'_{n+\frac{1}{2}}\left( 
\alpha_n\right)\right]^2}\; , \non
\end{eqnarray}
where $\sum_{\alpha_n}$ denotes the infinite sum over all positive zeros 
$\alpha_n$ of the Bessel function $J_{n+\frac{1}{2}}(\alpha_n)=0$, $P_n$ 
is the n-th Legendre polynomial and $\mu$ is the cosine of the angle between 
$\mb{r}$ and $\mb{r}_0$.\\
The density of finding the particle at arbitrary angles 
$\varphi,\,\,\vartheta$ at radius $r$ is then given by
\begin{eqnarray}
\rho_r(r,t|r_0,t_0)=\hspace*{5cm}\nnn
\int_0^{2\pi}d\varphi \int_{0}^\pi d\vartheta 
P_D(r,\varphi,\vartheta,t|r_0,\varphi_0,\vartheta_0,t_0)\,\sin(\vartheta)\, 
r^2\nnn
=\frac{2r^\frac{3}{2}}{R^2 \sqrt{r_0}} \sum_{\alpha_0} 
e^{-\alpha_0^2\frac{D(t-t_0)}{R^2}}\dfrac{ J_{\frac{1}{2}}\left( 
\alpha_0\frac{r_0}{R}\right)J_{\frac{1}{2}}\left( 
\alpha_0\frac{r}{R}\right)}{\left[J'_{\frac{1}{2}}\left( 
\alpha_0\right)\right]^2}\quad \nnn
=\frac{2r}{R r_0} \sum_{n=1}^\infty 
e^{-n^2\pi^2\frac{D(t-t_0)}{R^2}}\sin\left(\frac{n\pi 
r_0}{R}\right)\sin\left(\frac{n\pi r}{R}\right)\; .\non
\end{eqnarray}

The corresponding distribution function can be derived by integrating 
the $\sin$-functions:
\begin{eqnarray}
F_r(r,t|r_0,t_0)=\frac{2}{R r_0} \sum_{n=1}^\infty 
e^{-n^2\pi^2\frac{D(t-t_0)}{R^2}}\sin\left(\frac{n\pi 
r_0}{R}\right)\nnn
\cdot\left[\frac{R^2}{n^2\pi^2} \sin\left(\frac{n\pi 
r}{R}\right)-\frac{R}{n\pi}r\cos\left(\frac{n\pi r}{R}\right) \right]\non
\end{eqnarray}

Hence, the distribution function belonging to $\rho_b^D$ is given by
\begin{eqnarray}
F_b^D(t|r_0,t_0)=\hspace*{5cm}\nnn
1-\frac{2R}{\pi r_0}\sum_{n=1}^\infty 
e^{-n^2\pi^2\frac{D(t-t_0)}{R^2}}\sin\left(\frac{n\pi r_0}{R}\right) 
\frac{(-1)^{n+1}}{n}\; ,\nonumber
\end{eqnarray}
which was used to derive Eq. (\ref{assumption_correct_3d}) with the help 
of l'Hospital's rule ($r_0\rightarrow 0 $).\\
A distribution function for $\mu\in[-1;1]$ under the condition of being 
at radius $r$ can be derived straightforwardly by integrating the 
Legendre polynomials $P_n$. Using the sampled $\mu$, the angels $\varphi$ 
and $\vartheta$ are sampled.

\end{document}